\documentclass[hidelinks,conference,letterpaper]{IEEEtran}

\usepackage{bm}
\usepackage{amsmath}
\usepackage{extarrows}
\usepackage{amssymb}
\usepackage{graphicx}
\usepackage{color, xcolor}
\usepackage{enumerate}
\usepackage{bookmark}
\graphicspath{{figure}}
\usepackage{setspace}
\usepackage{multirow}
\usepackage{amsmath}
\usepackage{subfigure}
\usepackage{float}
\usepackage{booktabs}
\usepackage{algorithm}
\usepackage{algorithmic}
\usepackage{cite}
\usepackage{framed} 
\usepackage{scalerel}

\usepackage{tikz}
\usetikzlibrary{matrix}
\usepackage{lipsum}

\newcommand{\mr}{\mathrm}

\newcommand{\BE}{\begin{equation}}
\newcommand{\EE}{\end{equation}}
\newcommand{\BS}{\begin{subequations}}
\newcommand{\ES}{\end{subequations}}
\renewcommand{\bf}{\bm}

\newtheorem{theorem}{Theorem}

\newtheorem{assumption}{Assumption}
\newtheorem{definition}{Definition}

\newtheorem{remark}{Remark}
\newtheorem{lemma}{Lemma}
\newtheorem{corollary}{Corollary}

\DeclareMathOperator*{\argmin}{arg\,min}

\title{Sufficient Statistic Memory Approximate\\ Message Passing}
\author{\normalsize\IEEEauthorblockN{Lei~Liu, \emph{Member, IEEE}, Shunqi~Huang, and  Brian~M.~Kurkoski, \emph{Member, IEEE} \vspace{1mm}\\
  School of Information Science, Japan Institute of Science and Technology (JAIST), Nomi 923-1292, Japan}}

\date{ }

\begin{document}

\maketitle

\begin{abstract}
Approximate message passing (AMP) type algorithms have been widely used in the signal reconstruction of certain large random linear systems. A key feature of the AMP-type algorithms is that their dynamics can be correctly described by state evolution. However, state evolution does not necessarily guarantee the convergence of iterative algorithms. To solve the convergence problem of AMP-type algorithms in principle, this paper proposes a memory AMP (MAMP) under a sufficient statistic condition, named sufficient statistic MAMP (SS-MAMP). We show that the covariance matrices of SS-MAMP are L-banded and convergent. Given an arbitrary MAMP, we can construct the SS-MAMP by damping, which not only ensures the convergence, but also preserves the orthogonality, i.e., its dynamics can be correctly described by state evolution.  
\end{abstract}

\textit{A full version of this paper is accessible at
\href{https://arxiv.org/pdf/2112.15327.pdf}{arXiv} (see \cite{Lei2022SSMAMP}).} 

\section{Introduction}
This paper studies the signal recovery of noisy linear systems: $\bf{y}=\bf{Ax}+\bf{n}$, where $\bf{y}\in \mathbb{C}^{M\times1}$ is the observed vector, $\bf{A}\in \mathbb{C}^{M\times N}$ a known matrix, $\bf{x}\in \mathbb{C}^{N\times1}$ the signal vector to be recovered, and $\bf{n}\in \mathbb{C}^{M\times1}$ a Gaussian noise vector. The goal is to recover $\bf{x}$ using  $\bf{y}$ and $\bf{A}$ . This paper focuses on large-size systems with $M,N\to\infty$ and a fixed $\delta=M/N$. For non-Gaussian $\bf{x}$, the optimal recovery is in general NP-hard \cite{Micciancio2001,verdu1984_1}.

\subsection{Background}
 Approximate message passing (AMP) is a fantastic approach for the signal recovery of high-dimensional noisy linear systems  \cite{Donoho2009}. AMP has three crucial properties. First, the complexity of AMP is very low as it suppresses the linear interference using a simple matched filter. Second, the dynamics of AMP can be tracked by state evolution (SE) \cite{Bayati2011}. Most importantly, it was proved that AMP is  minimum mean square error (MMSE) optimal for un-coded signaling \cite{Barbier2017arxiv, Reeves_TIT2019}  and information-theoretically (i.e., capacity) optimal for coded signaling \cite{LeiTIT}. However, AMP is restricted to independent and identically distributed (IID) Gaussian $\bf{A}$ \cite{Rangan2015, Vila2014}, which limits the application of AMP. To solve the weakness of AMP, a unitarily-transformed AMP (UTAMP) was proposed in \cite{UTAMP}, which works well for correlated matrices $\bf{A}$. Orthogonal/vector AMP (OAMP/VAMP) was proposed for right-unitarily-invariant matrices \cite{Ma2016,Rangan2016}.  The SE of OAMP/VAMP was conjectured in \cite{Ma2016} and proved in \cite{Rangan2016,Takeuchi2017}. The MMSE (predicted by replica method) optimality and the information-theoretical optimality of OAMP/VAMP were proved in \cite{Ma2016, Tulino2013, Barbier2018b,Kabashima2006} and \cite{Lei_cap_oamp}, respectively, when the SE has a unique fixed point. However, OAMP/VAMP requires high-complexity linear MMSE (LMMSE) estimation, which is prohibited in practical large-scale systems.

To solve the weakness of OAMP/VAMP, a low-complexity convolutional AMP (CAMP) was proposed in  \cite{Takeuchi2020CAMP}. However, CAMP may fail to converge, especially for ill-conditioned matrices. Apart from that, the AMP  was extended to solve the Thouless-Anderson-Palmer (TAP) equations in Ising models for unitarily-invariant matrices \cite{Opper2016}. The results in \cite{Opper2016} were justified via SE in \cite{Fan2020arxiv}. More recently, memory AMP (MAMP) was established for unitarily-invariant matrices \cite{Lei2020MAMPTIT, Lei2020MAMPISIT}. MAMP has a comparable complexity to AMP since it use a low-complexity memory matched filter to suppresses the linear interference. In addition, the dynamics of MAMP can be correctly predicted by SE. More importantly, the convergence of MAMP is guaranteed by a long-memory optimized damping. Apart from that, MAMP achieves the MMSE (predicted by replica method) when its SE has a unique fixed point.

\subsection{Motivation and Related Works}  
A key feature of the AMP-type algorithms \cite{Donoho2009,Ma2016,Rangan2016,Takeuchi2020CAMP,Lei2020MAMPTIT, Lei2020MAMPISIT, Opper2016} is that their dynamics can be rigorously described by state evolution \cite{Rangan2016,Takeuchi2017,Bayati2011, Fan2020arxiv}. However,  state evolution does not necessarily guarantee the convergence of iterative algorithms \cite{Takeuchi2021OAMP, Lei2016TWC,Lei2019TWC,Gerbelot,RamjiPCA}. Therefore, it is desired to find a new technique or framework that ensures the convergence of the AMP-type algorithms.

Damping is generally used to improve the convergence of the AMP-type algorithms for finite-size systems \cite{Schniter2019dAMP,  Schniter2021dvamp, Takeuchi2019damp}. However, improper damping may break the orthogonality and Gaussianity, which may result in that the dynamics of damped AMP-type algorithms cannot be described by state evolution anymore \cite{Takeuchi2020CAMP}. Moreover,  the conventional damping in the literature was heuristic and empirical, and performed on the messages in current and last iterations. In \cite{Lei2020MAMPTIT, Lei2020MAMPISIT}, the authors first proposed an analytically optimized vector damping for Bayes-optimal MAMP (BO-MAMP) based on the current and all the preceding messages, which not only solves the convergence problem of MAMP, but also preserve the orthogonality and Gaussianity, i.e., the dynamics of damped BO-MAMP can be correctly described by state evolution. Recently, the damping optimization in \cite{Lei2020MAMPTIT, Lei2020MAMPISIT} was used to analyze the convergence of Bayes-optimal OAMP/VAMP in \cite{Takeuchi2021OAMP} from a sufficient statistic perspective. The works in \cite{Lei2020MAMPTIT, Lei2020MAMPISIT, Takeuchi2021OAMP} pave the way for a novel principle to solve the convergence of the AMP-type algorithms.

\subsection{Contributions}  
Motivated by the damping optimization in \cite{Lei2020MAMPTIT, Lei2020MAMPISIT} and its statistical interpretation in \cite{Takeuchi2021OAMP}, this paper proposes a sufficient statistic memory AMP (SS-MAMP) to solve the convergence problem of AMP-type algorithms in principle. SS-MAMP is an MAMP \cite{Lei2020MAMPTIT, Lei2020MAMPISIT} under a sufficient statistic condition. That is, from the MMSE perspective, the current output of each local processor is a sufficient statistic of the current and all preceding messages. We show that SS-MAMP has the following interesting properties.
\begin{itemize}
    \item SS-MAMP can be equivalently defined as an MAMP with L-banded covariance matrices (see Definition \ref{Def:L-baned}) whose elements in each ``L band" are the same. As a result, the original state evolution  on covariance matrices is reduced to a state evolution  on the diagonal variances. Furthermore, some interesting properties are derived for the inverse of the L-banded covariance matrices.
    
    \item The covariance matrices in SS-MAMP are monotonically decreasing and converge respectively to a certain value with the increase of the number of iterations. Hence, the state evolution of SS-MAMP is definitely convergent.
    
    \item Damping is useless, i.e., damping does not bring MSE improvement in SS-MAMP. 
    
    \item For the Bayes-optimal local processor in SS-MAMP, memory is useless, i.e., jointly estimation with preceding messages does not bring MSE improvement to the Bayes-optimal local processor in SS-MAMP.  
\end{itemize}

Given an arbitrary MAMP, we can construct an SS-MAMP using the optimal damping in \cite{Lei2020MAMPTIT, Lei2020MAMPISIT}. The constructed SS-MAMP has the following interesting properties.
\begin{itemize}
    \item The sufficient statistic property (i.e.,  L-banded covariance matrices) is guaranteed by the optimal damping at each local processor. Therefore, the constructed SS-MAMP inherits all the properties of SS-MAMP.
    
    \item  The orthogonality and Gaussianity of the original MAMP are preserved in the constructed SS-MAMP. Hence, its dynamics can be rigorously described by state evolution. 
    
    \item The MSE of the constructed SS-MAMP with optimal damping is not worse than the original MAMP.
\end{itemize}  
 
\subsection{Notation}   
  $\langle\bf{A}_{M\times N}, \bf{B}_{M\times N} \rangle \equiv \bf{A}^{\rm H}_{M\times N}\bf{B}_{M\times N}$,  $\mr{var}\{\bf{x}|\bf{Y}\} = \tfrac{1}{N}{\rm E}\big\{ \|\bf{x}- {\rm E}\{\bf{x}|\bf{Y}\} \|^2 \big\}$. A matrix is said column-wise IID Gaussian and row-wise joint-Gaussian if its each column is IID Gaussian and its each row is joint Gaussian.

\section{Preliminaries}\label{Sec:Pre}
We first give the problem formulation and assumptions. We then briefly introduce MAMP and its properties.
\subsection{Problem Formulation} 
We rewrite the noisy system with two constraints: 
\BS\label{Eqn:unitary_sys}\begin{align}
&\Gamma: \quad  \bf{y}=\bf{Ax}+\bf{n},\\
&\Phi: \quad x_i\sim P_x, \;\;\forall i.
\end{align}\ES 
where $\bf{n}\sim \mathcal{CN}(\bf{0},\sigma^2\bf{I})$, $\bf{x}$ is a length-$N$ IID vector following a distribution $P_{x}$, and $\bf{A}$ is an $M\times N$ measurement matrix. In addition, we assume that $M,N\to\infty$ with a fixed $\delta=M/N$.

\begin{assumption}\label{ASS:x}
The entries of $\bf{x}$ are IID with zero mean. The variance of $\bf{x}$ is normalized, i.e., $\frac{1}{N}{\rm E}\{\|\bf{x}\|^2\}\overset{\rm a.s.}{=}1$, and the $(2+\epsilon)$th moments of  $\bf{x}$ are finite for some $\epsilon$. 
\end{assumption}
 
 \begin{assumption}\label{ASS:A}
  Let the singular value decomposition of $\bm{A}$ be $\bf{A}\!=\!\bf{U}\bf{\Sigma} \bf{V}^{\rm H}$, where $\bf{U}\!\in\! \mathbb{C}^{M\times M}$ and $\bf{V}\!\in\! \mathbb{C}^{N\times N}$ are unitary matrices, and $\bf{\Sigma} $ is a rectangular diagonal matrix.  We assume that $M,\!N\!\to\!\infty$ with a fixed $\delta\!=\!\!M\!/\!N$, and $\bf{A}$ is known and is right-unitarily-invariant, i.e., $\bf{U}\bf{\Sigma}$ and $\bf{V}$ are independent, and $\bf{V}$ is Haar distributed (or equivalently, isotropically random). Furthermore, the empirical eigenvalue distribution of $\bf{AA}^{\rm H}$ converges almost surely to a compactly supported deterministic distribution with unit first moment in
the large system limit, i.e., $ \tfrac{1}{N} {\rm tr}\{\bf{A}\bf{A}^{\rm H}\}\overset{\rm a.s.}{=}1$  \cite{Takeuchi2020CAMP}.  
\end{assumption}\vspace{-2mm}


 \subsection{Memory AMP}\label{Sec:MIP}
\begin{definition}[Memory  Iterative Process]\label{Def:MIP}
 A memory iterative process consists of a memory linear estimator (MLE) and a memory non-linear estimator (MNLE) defined as:  Starting with $t=1$, 
\BS\label{Eqn:MIP}\begin{alignat}{2}
{\rm MLE:}&& \quad \quad  \bf{r}_t &= \gamma_t\!\left(\bf{X}_{t}\right)=  \bf{\mathcal{Q}}_t\bf{y} + \textstyle\sum_{i=1}^t{\bf{\mathcal{P}}}_{t,i} \bf{x}_i,\label{Eqn:MIP_LE}\\
{\rm MNLE:}&& \quad   \bf{x}_{t + 1} &= \phi_t\!\left(\bf{R}_t\right).
\end{alignat}\ES 
 where $\bf{X}_t=[\bf{x}_1\dots\,\bf{x}_t]$, $\bf{R}_t=[\bf{r}_1\dots\,\bf{r}_t]$, $\bf{\mathcal{Q}}_t\bf{A}$ and $\bf{\mathcal{P}}_{t,i}$ are polynomials in $\bf{A}^{\rm H}\bf{A}$. Without loss of generality, we assume that the norms of $\bf{\mathcal{Q}}_t$ and $\{\bf{\mathcal{P}}_{t,i}\}$ are finite. Hence, $\gamma_t\!\left(\cdot\right)$ in \eqref{Eqn:MIP_LE} is Lipschitz-continuous \cite{Lei2020MAMPISIT, Lei2020MAMPTIT}.
\end{definition} 

The memory  iterative process is degraded to the conventional non-memory iterative process if $ \gamma_t\!\left(\bf{X}_t\right)= {\gamma}_t \left(\bf{x}_t\right) $ and $\phi_t\!\left(\bf{R}_t\right)={{\phi}}_t \left(\bf{r}_t\right)$. Let $ \bf{R}_t  = \bf{X} + \bf{G}_t$ and $\bf{X}_t = \bf{X} + \bf{F}_t$, where $\bf{X}=\bf{x}\cdot\bf{1}^{\rm T}$, $\bf{1}$ is an all-ones vector with proper size,  and  $\bf{G}_t=[\bf{g}_1\dots\,\bf{g}_t]$ and $\bf{F}_t=[\bf{f}_1\dots\,\bf{f}_t]$ indicate the estimation errors with zero means and the covariances as follows:
\BE\label{Eqn:v_gamma_phi} 
    \bf{V}_t^{\gamma}   \equiv \tfrac{1}{N} \langle \bf{G}_t,  \bf{G}_{t}\rangle, \qquad 
   \bf{V}_t^{\phi} \equiv \tfrac{1}{N} \langle \bf{F}_t, \bf{F}_{t} \rangle. 
\EE 
Let $\{v^{\gamma}_{i,j}\}$ and $\{v^{\phi}_{i,j}\}$ be the elements of $ \bf{V}_t^{\gamma}$ and  $\bf{V}_t^{\phi}$, respectively. Let $v^{\gamma}_{i}=v^{\gamma}_{i,i}$ and  $v^{\phi}_{i}=v^{\phi}_{i,i}$. We define the diagonal vectors of the covariance matrices as
\BE\label{Eqn:vars}
  \bf{v}_t^{\gamma} \equiv[v^{\gamma}_{1}\dots\,  v^{\gamma}_{t}]^{\rm T}, \quad 
  \bf{v}_t^{\phi} \equiv[v^{\phi}_{1}\dots\,  v^{\phi}_{t}]^{\rm T}.
\EE

\begin{definition}[Memory AMP \cite{Lei2020MAMPISIT, Lei2020MAMPTIT}]\label{Def:MAMP}
The memory  iterative process is said to be a memory AMP (MAMP) if the following orthogonal constraint holds for $t\ge 1$:
\BE\label{Eqn:Orth_MAMP} 
 \tfrac{1}{N} \langle \bf{g}_t, \bf{x} \rangle  \overset{\rm a.s.}{=}  0, \quad\!\!
    \tfrac{1}{N} \langle \bf{g}_t, \bf{F}_t\rangle    \overset{\rm a.s.}{=}  \bf{0},\quad\!\!
 \tfrac{1}{N} \langle \bf{f}_{t+1}, \bf{G}_t \rangle      \overset{\rm a.s.}{=} \bf{0}. 
  \EE      
\end{definition}
 
The lemma below shows the  IID Gaussianity of MAMP.
 
\begin{lemma}\label{Lem:IIDG_MIP}
 Suppose that Assumptions 1-2 hold. For MAMP with the orthogonality in \eqref{Eqn:Orth_MAMP},  the ${\gamma}_t$ in \eqref{Eqn:MIP_LE} and separable-and-Lipschitz-continuous  $\phi_t$  \cite{Berthier2020},  we have \cite{Takeuchi2020CAMP}: $\forall 1\!\!\le \!\!t'\!\!\leq\!\! t$, \vspace{-4mm}
\BS\label{Eqn:IIDG_MIP}  \begin{align}
v_{t,t'}^{\gamma} &\!\!\overset{\rm a.s.}{=}\!\! \tfrac{1}{N} \langle \gamma_t\big(\bf{X}+\bf{N}_{t}^{\phi}\big)-\bf{x}, \gamma_{t'}\big(\bf{X}+\bf{N}_{t'}^{\phi}\big)-\bf{x}\rangle,\\ 
v_{t+1,{t'}\!+1}^{\phi} &\!\!\overset{\rm a.s.}{=} \!\!\tfrac{1}{N} \langle \phi_t\big(\bf{X}+\bf{N}_{t}^{\gamma} \big)\!-\!\bf{x}, \phi_{t'}\big(\bf{X}+\bf{N}_{t'}^{\gamma} \big)-\bf{x}\rangle,  
\end{align}\ES 
where $\bf{N}_{t}^{\gamma}=[\bf{n}_{1}^{\gamma}\dots\, \bf{n}_{t}^{\gamma}]$ and  $\bf{N}_{t}^{\phi}=[\bf{n}_{1}^{\phi} \dots\, \bf{n}_{t}^{\phi}]$ are  column-wise IID Gaussian, row-wise joint-Gaussian and independent of $\bf{x}$. In addition, $\tfrac{1}{N}\langle \bf{N}_{t}^{\gamma}, \bf{N}_{t}^{\gamma} \rangle =\bf{V}_{t}^{\gamma}$ and $ \tfrac{1}{N} \langle \bf{N}_{t}^{\phi}, \bf{N}_{t}^{\phi} \rangle =\bf{V}_{t}^{\phi}$. In detail, $\bf{n}_t^{\phi}\!\!\sim\!\mathcal{CN}(\bf{0},v_{t,t}^{\phi}\bf{I})$ with ${\rm E}\{\bf{n}_t^{\phi}(\bf{n}_{t'}^{\phi})^{\rm H}\}\!=\!v_{t,t'}^{\phi}\bf{I}$ and $\bf{n}_t^{\gamma}\!\!\sim\!\mathcal{CN}(\bf{0},v_{t,t}^{\gamma}\bf{I})$ with ${\rm E}\{\bf{n}_t^{\gamma}(\bf{n}_{t'}^{\gamma})^{\rm H}\}\!=\!v_{t,t'}^{\gamma}\bf{I}$. 
\end{lemma}

The error covariance matrices of MAMP can be tracked by the following state evolution: Starting with $v_{1,1}^{\phi}=1$,
\BS\label{SE_MAMP}\begin{align} 
{\rm MLE:} \quad \quad\bf{V}_t^{\gamma}&= \gamma_{\mr{SE}}(\bf{V}_t^{{\phi}}),\label{SE_MAMP_MLE}\\
{\rm NLE:} \; \quad  \bf{V}_{t+1}^{{\phi}}  &= {\phi}_{\mr{SE}}(\bf{V}_t^{\gamma}),\label{SE_MAMP_NLE}    
\end{align}\ES
where $\gamma_{\mr{SE}}(\cdot)$ and ${\phi}_{\mr{SE}}(\cdot)$ are the MSE transfer functions corresponds to $\gamma_t(\cdot)$ and $\phi_t(\cdot)$, respectively.  
 
 \section{Sufficient Statistic and L-Banded Matrices}\label{Sec:SS-L-banded}
  We introduce the sufficient statistic, L-banded matrix, and their properties that will be used in this paper.

 \subsection{Sufficient Statistic}
 The following lemma provides a sufficient statistic property of Gaussian observations. 
  
 \begin{lemma}[Sufficient Statistic of Gaussian Observations]\label{Lem:SS_Est}
Let $\bf{x}_t = x\bf{1}_{t\times 1}  + \bf{n}_t$, where $\bf{x}_t = [x_{1} \dots x_{t}]^{\rm T}$, $\bf{n}_t\sim \mathcal{CN}(\bf{0},\bf{V}_t)$, and $\bf{V}_t$ is invertible. For any $\bf{x}_t$, $x_{t}$ is a sufficient statistic of $\bf{x}_t$, i.e., $\bf{x}_{t-1}$ --- $x_t$ --- $x$ is a Markov chain or
  \begin{align} 
    p(x|\bf{x}_t) &=p(x|x_t),\label{Eqn:SS_p}
\end{align}     
  if and only if  
\BE
    v_{t,i}=v_{i,t}=v_{t,t}, \quad \forall 1\le i\le t,\label{Eqn:ss_v}
\EE
i.e., the elements in last row and last column of $\bf{V}_t$ are the same. If $x_{t}$ is a sufficient statistic of $\bf{x}_t$, damping is then useless for $\bf{x}_t$ (i.e., does not improve the MSE), i.e.,  
\BE\label{Eqn:useless_dam}
 [0\dots 0\;1]^{\rm T} = \argmin_{\bf{\zeta}_{t}:\bf{\zeta}_{t}^{\rm T}\bf{1}=1} \tfrac{1}{N}\|\bf{\zeta}_{t}^{\rm T}\bf{x}_t -{x} \|^2.
\EE
\end{lemma} 
  
 \begin{IEEEproof}
 See Appendix \ref{APP:SS_Est}.  
 \end{IEEEproof}  
 
  The following is a  corollary of Lemma \ref{Lem:SS_Est}. 
 
 \begin{corollary}\label{Cor:SS_Est}
Let $\bf{x}_t = x\bf{1}_{t\times 1}  + \bf{n}_t$, where $\bf{x}_t = [x_{1} \dots x_{t}]^{\rm T}$, $\bf{n}_t\sim \mathcal{CN}(\bf{0},\bf{V}_t)$, and $\bf{V}_t$ is invertible. We assume that  $x_{t}$ is a sufficient statistic of $\bf{x}_t$, i.e., \eqref{Eqn:SS_p} holds for all $\bf{x}_t$. Then, 
\BE
    {\rm E}\{x|\bf{x}_t\}  ={\rm E}\{x|x_t\},\label{Eqn:SS_E} 
\EE
Furthermore, if $y$ --- $x$ --- $\bf{x}_t$ is a Markov chain, for any $\{y, \bf{x}_t\}$, $\{y, x_{t}\}$ is a sufficient statistic of $\{y, \bf{x}_t\}$, i.e.,
\BS \begin{align} 
    p(x|y, \bf{x}_t) &=p(x|y, x_t),\label{Eqn:SS_py}\\
    {\rm E}\{x|y,\bf{x}_t\} &={\rm E}\{x|y,x_t\}.\label{Eqn:SS_Ey} 
\end{align} \ES 
 \end{corollary}
  
 Corollary \ref{Cor:SS_Est} shows that if $x_t$ is a sufficient statistic of Gaussian observations $\bf{x}_t$, the \emph{a-posterior} (i.e., MMSE) estimation of $x$ based on $\bf{x}_t$ is equivalent to that based on $x_t$.
 
 The following lemma gives the construction of a sufficient statistic of the Gaussian observations.
 
  \begin{lemma} \label{Lem:SS_damp}
Let $\bf{x}_t = x\bf{1}_{t\times 1}  + \bf{n}_t$, where $\bf{x}_t = [x_{1} \dots x_{t}]^{\rm T}$, $\bf{n}_t\sim \mathcal{CN}(\bf{0},\bf{V}_t)$, and $\bf{V}_t$ is invertible. Then, a sufficient statistic of $\bf{x}_t$ can be constructed by optimal damping \cite{Lei2020MAMPISIT, Lei2020MAMPTIT}:
\BS\label{Eqn:damp}\BE
   x_{t+1} = \bf{x}_t\bf{\zeta}_t^{\rm T},
\EE
where 
\BE
    \bf{\zeta}_t =  {\bf{V}_t^{-1}\bf{1}}/[{\bf{1}^{\rm T}\bf{V}_t^{-1}\bf{1}}].
\EE\ES 
\end{lemma} 
 
 \begin{lemma}\label{Lem:decreas_V}
  Suppose that $\bf{V}_t$ is  a $t\times t$ covariance matrix. If the elements in last row and last column of $\bf{V}_t$ are the same, i.e.,  \vspace{-2mm}
\BE
    v_{t,i}=v_{i,t}=v_{t,t}, \quad \forall 1\le i\le t,
\EE
we then have \vspace{-2mm}
 \BS\begin{align}
   v_{t,t}  &\le v_{i,i},\quad \forall 1\le i\le t.
 \end{align} \ES 
\end{lemma}

\begin{IEEEproof}
   Following the covariance inequality,  we have $v_{t,i} \le \sqrt{v_{t,t} v_{i,i}}$.  Since $v_{t,i} =v_{t,t}, \forall i\le t$, we then have $v_{t,t} \le  v_{i,i}, \forall i \le t$.  
   \end{IEEEproof}

 \subsection{L-Banded  Matrix} 
 \begin{definition}[L-Banded Matrix]\label{Def:L-baned}
 An $t\times t$  matrix $\bf{V}=\{v_{i,j}\}$ is said to be L-banded if  the elements in each ``L band" of the matrix are the same, i.e.,
\BE
   v_{i,j}  =  v_{\max(i,j)}, \quad \forall 1\le i\le t, \forall 1\le j\le t. \label{Eqn:SS_cov}  
\EE  
That is,
 \BE \footnotesize
   \bf{V}= \begin{tikzpicture}[baseline=-\dimexpr-.0mm\relax]
  \matrix [matrix of math nodes,left delimiter=(,
 right delimiter=), row sep=3mm,column sep=3mm, ampersand replacement=\&] (M) {
   v_1 \& v_2 \& \cdots \& v_t          \\ 
   v_2 \& v_2 \& \cdots \& v_t          \\ 
   \vdots \& \vdots \& \ddots \& \vdots   \\
   v_t \& v_t \& \cdots  \& v_t             \\ 
   };
   \draw[](M-1-2.north east)--(M-2-2.south east)--(M-2-1.south west)--(M-2-1.north west)--(M-2-2.north west)--(M-1-2.north west)--(M-1-2.north east);
   \draw[](M-1-4.north east)--(M-4-4.south east)--(M-4-1.south west)--(M-4-1.north west)--(M-4-4.north west)--(M-1-4.north west)--(M-1-4.north east);
\end{tikzpicture}.
 \EE 
\end{definition}

 \begin{lemma}[Inverse of L-Banded  Matrix]\label{Lem:SS_matrix_inv}
 Suppose that $\bf{V}$ is  an $t\times t$ invertible L-banded  matrix with diagonal elements $\{v_1,\dots, v_t\}$, and $\bf{V}^{-1}=\{v_{i,j}^\dagger\}$ be the inverse of $\bf{V}$. Let $v_0\equiv \infty$, $v_{t+1}\equiv0$ and $\delta_i={v_{i}-v_{i+1}}$. Then, $\bf{V}^{-1}$ is a tridiagonal matrix given by  
 \BE\label{Eqn:SS_V_inv}
    v_{i,j}^\dagger = \left\{ \begin{array}{ll} 
        \delta_{i-1}^{-1} + \delta_i^{-1}, & {\rm if}\;  1\le i=j\le t \vspace{1mm}\\ 
        -\delta^{-1}_{\min(i,j)}, & {\rm if}\;  |i-j|=1 \vspace{1mm}\\
        0, & {\rm otherwise}
    \end{array}\right..
 \EE 
  Furthermore, 
 \BS\label{Eqn:SSV_inv_sum}\begin{align}
     \bf{1}^{\rm T} \bf{V}^{-1}& = [ \bf{V}^{-1}\bf{1}]^{\rm T}  = \big[0, \dots, 0, v_t^{-1}\big],\\
     \bf{1}^{\rm T}\bf{V}^{-1}\bf{1} & = v_t^{-1},
 \end{align} \ES 
and $\bf{V}$ is invertible if and only if $v_i\neq v_j$ when $i\neq j$. 
\end{lemma}
 
 \begin{IEEEproof}
  It is easy to verify that $\bf{V}^{-1}\bf{V}=\bf{I}$. Hence, $\bf{V}^{-1}$ in \eqref{Eqn:SS_V_inv} is the inverse of L-banded covariance matrix $\bf{V}$.  Furthermore, \eqref{Eqn:SSV_inv_sum} straightforwardly follows \eqref{Eqn:SS_V_inv}. 
 \end{IEEEproof}
 
The following lemma gives the monotonicity and convergence of L-banded covariance matrices.
  
\begin{lemma}\label{Lem:converg_SS_V}
  Suppose that $\bf{V}$ is  an $t\times t$  L-banded covariance matrix with diagonal elements $\{v_1,\dots, v_t\}$. Then, the sequence $(v_1,\dots, v_t)$ is monotonically decreasing and converge to a certain value, i.e.,
 \BS\begin{align}
   v_i  &\le v_j,\quad {\rm if} \; j<i\le t,\\
   v_t  &\to v_*,\quad {\rm if} \;t\to\infty. 
 \end{align} \ES 
\end{lemma}

\begin{IEEEproof} 
    Since $\bf{V}$ is L-banded, following Lemma \ref{Lem:decreas_V}, we have $v_{t} \le  v_{t'}, \; \forall t' \le t$. 
    That is, $\{v_t\}$ is monotonically decreasing with $t$. Furthermore, $v_{t}>0, \forall t \le t$, i.e., sequence $(v_1,\dots, v_t)$ has a lower bound 0. Hence, if $t\to\infty$, sequence $\{v_{t}\}$ converges to a certain value, i.e., $\lim_{t\to\infty} v_t \to v_*$. Therefore, we complete the proof of Lemma \ref{Lem:converg_SS_V}. 
   \end{IEEEproof}

Give an arbitrary sequence, the following lemma constructs a new sequence with an L-banded covariance matrix.

 \begin{lemma}\label{Lem:L-banded_seq}
Let $\bar{\bf{x}}_t = x\bf{1}_{t\times 1}  + \tilde{\bf{n}}_t$, where  $\tilde{\bf{n}}_t$ is zero mean with covariance matrix $\tilde{\bf{V}}_t$, which is not necessarily L-banded. Assume that $\tilde{v}_{i,i}>0, 1\le i\le t$.  We can construct a new sequence: For $1\le i\le t$,
\BS\label{Eqn:damp_seq}\BE
    x_i = \tilde{\bf{x}}_i^{\rm T} \bf{\zeta}_i,  
\EE
where 
\BE
    \bf{\zeta}_i =  \left\{\!\!\!\begin{array}{ll}
              \frac{  \tilde{\bf{V}}_{i}^{-1} \bf{1}}{\bf{1}^{\rm T} \tilde{\bf{V}}_{i}^{-1}\bf{1}}, &\;\; {\rm if}\;   \tilde{\bf{V}}_{i} {\rm \;is\; invertible} \vspace{1mm}\\
               {\rm [}\bf{\zeta}_{i-1}^{\rm T} \; 0{\rm ]}^{\rm T} , & \;\; {\rm otherwise}
        \end{array}\right.. 
\EE\ES
In addition, if $\tilde{\bf{V}}_{i}$ is singular, $x_i=x_{i-1}$. In this case, we will not allow $\tilde{x}_{i}$ to join the damping for $\{x_j, j>i\}$. Then, we have ${\bf{x}}_t = x\bf{1}_{t\times 1}  + {\bf{n}}_t$, where  ${\bf{n}}_t$ is zero mean with L-banded covariance matrix ${\bf{V}}_t$.
\end{lemma} 

Intuitively, the $i$-th damping $x_i = \tilde{\bf{x}}_i^{\rm T} \bf{\zeta}_i$ in \eqref{Eqn:damp_seq} ensures that the elements in the $i$-th band of $\bf{V}_t$ are the same. Hence,  the step-by-step damping in \eqref{Eqn:damp_seq} results in an L-banded $\bf{V}_t$.

\section{Sufficient Statistic MAMP (SS-MAMP)}\label{Sec:SS-MAMP}
  Even though it is correct, the state evolution of MAMP may not converge or even diverge if it is not properly designed. In this section, using the sufficient statistic technique in \ref{Sec:SS-L-banded}, we construct an SS-MAMP to solve the convergence problem of MAMP in principle.

 \subsection{Definition and Properties of SS-MAMP} 

\begin{definition}[Sufficient Statistic MAMP]\label{def:ss_mamp}
 An MAMP is said to be sufficient statistic if for all $t$,
 \BS\label{Eqn:SS_mmse}\begin{align} 
    \mr{var}\{\bf{x}|\bf{R}_t\}=\mr{var}\{\bf{x}|\bf{r}_t\},\\
    \mr{var}\{\bf{x}|\bf{y},\bf{X}_t\} = \mr{var}\{\bf{x}|\bf{y},\bf{x}_t\}.
\end{align} \ES    
That is,  in SS-MAMP, $\bf{r}_t$ and $\bf{x}_t$ are respectively sufficient statistic of $\bf{R}_t$ and $\bf{X}_t$ from the MMSE perspective.   
\end{definition}

The following lemma follows Definition \ref{def:ss_mamp}.

\begin{lemma}  
Suppose that Assumptions 1-2 hold. For the Bayes-optimal local processor (e.g., \emph{a-posterior} estimator) in SS-MAMP, memory is useless, i.e., jointly estimation with preceding messages does not bring improvement in MSE. 
\end{lemma}

Following Lemmas \ref{Lem:IIDG_MIP} and \ref{Lem:SS_Est}, we give an equivalent definition of SS-MAMP in the following.

\begin{lemma}[Sufficient Statistic of SS-MAMP]\label{Lem:SS} 
Suppose that Assumptions 1-2 hold. Then, an MAMP is  sufficient statistic if and only if the covariance matrices $\bf{V}^{\gamma}_t$ (of $\bf{R}_t$) and $\bf{V}^{\phi}_t$  (of $\bf{X}_t$) are L-banded, i.e., for all $t$, $1\le \{i, j\}\le t$,
\BE\label{Eqn:SS} 
     v^{\gamma}_{i,j}  =  v^{\gamma}_{\max(i,j)}, \qquad  
   v^{{\phi}}_{i,j} = v^{{\phi}}_{\max(i,j)}.  
  \EE   
\end{lemma}

In detail, following Lemma \ref{Lem:SS_Est}, in $i$-th iteration, the sufficient statistic of $r_i$ over $\bf{r}_i$ (or $x_i$ over $\bf{x}_i$) is equivalent to the condition that the elements in the last row and last column of $\bf{V}_i^\gamma$ (or $\bf{V}_i^\phi$) are the same. Since SS-MAMP requires the sufficient statistic of $r_t$ (or $x_t$) for all iterations, it is equivalent to the condition that for all $t$, the elements in the last row and last column of $\bf{V}_t^\gamma$ (or $\bf{V}_t^\phi$) are the same.

\begin{lemma}[State Evolution of SS-MAMP]\label{Lem:SE}
Suppose that Assumptions 1-2 hold. In SS-MAMP, the covariance matrices $\bf{V}_t^{\gamma}$ and $\bf{V}_t^{\phi}$ are determined by their  diagonal sequences $  \bf{v}_t^{\gamma} =[v^{\gamma}_{1}\dots\,  v^{\gamma}_{t}]^{\rm T}$ and $\bf{v}_t^{\phi} =[v^{\phi}_{1}\dots\,  v^{\phi}_{t}]^{\rm T}$, respectively. The state evolution of SS-MAMP can be simplified to: Starting with $t=1$ and $v^\phi_{1}=1$,
     \begin{align}
     v^\gamma_{t}  =\gamma_{\mr{SE}}(\bf{v}^\phi_t),  \quad
      {v}^\phi_{t+1}  =  \phi_{\mr{SE}}(\bf{v}^\gamma_t), \label{Eqn:SE_MAMP_NLE}
    \end{align} 
    where $\gamma_{\mr{SE}}(\cdot)$ and $\phi_{\mr{SE}}(\cdot)$ are the MSE transfer functions corresponds to $\gamma_t(\cdot)$ and $\phi_t(\cdot)$, respectively.
\end{lemma}

From Lemmas \ref{Lem:converg_SS_V} and  \ref{Lem:SS}, we have the following.

\begin{lemma}[Convergence of SS-MAMP] \label{Lem:converge}
 The diagonal sequences $(v^{{\gamma}}_1,\dots,v^{{\gamma}}_{t})$ and $(v^{{\phi}}_1,\dots,v^{{\phi}}_{t})$ in SS-MAMP are monotonically decreasing and converge respectively to a certain value, i.e., 
   \BS\begin{align}
   v^{{\gamma}}_{t} &\le v^{{\gamma}}_{t'}, \;  \forall t' \le t, &
        \lim\limits_{t\to\infty} v^{{\gamma}}_{t} &\to v^{{\gamma}}_*, \label{Eqn:mon_v_g}\\
       v^{{\phi}}_{t} &\le v^{{\phi}}_{t'}, \;  \forall t' \le t, &
        \lim\limits_{t\to\infty} v^{{\phi}}_{t} &\to v^{{\phi}}_*.\label{Eqn:mon_v_f}
   \end{align}\ES
   Therefore, the state evolution of SS-MAMP is  convergent.  
\end{lemma}

From Lemmas \ref{Lem:SS_Est} and \ref{Lem:SS}, we have the following.

\begin{lemma}[Damping is Useless in SS-MAMP]\label{Lem:damp_useless}
 Damping is useless (i.e., has no MSE improvement) in SS-MAMP. Specifically,  the optimal damping of $\bf{R}_t$ (or $\bf{X}_t$) has the same MSE as that of $\bf{r}_t$ (or $\bf{x}_t$).
\end{lemma}\vspace{-3mm}

\subsection{SS-MAMP Construction}
Given an arbitrary MAMP, the following constructs an SS-MAMP using damping.

 \begin{theorem}[SS-MAMP Construction]\label{The:SS-MAMP_con}
Suppose that Assumptions 1-2 hold. Given an arbitrary MAMP: 
\BS\label{Eqn:org_MAMP}\begin{align} 
  \tilde{\bf{r}}_t &= \gamma_t\!\left(\bf{X}_{t}\right),\label{Eqn:MAMP_g}\\
  \tilde{\bf{x}}_{t + 1} &= \phi_t\!\left(\bf{R}_t\right),\label{Eqn:MAMP_f}
\end{align}\ES  
where $\gamma_t(\cdot)$ is given in \eqref{Eqn:MIP_LE}, $\phi_t(\cdot)$ is a separable-and-Lipschitz-continuous, and the orthogonality in \eqref{Eqn:Orth_MAMP} holds. Let $\tilde{\bf{R}}_{t}=[\tilde{\bf{r}}_{1}\dots \tilde{\bf{r}}_{t}]$ and $\tilde{\bf{X}}_{t+1}=[\tilde{\bf{x}}_{1}\dots \tilde{\bf{x}}_{t+1}]$. Then, based on  \eqref{Eqn:org_MAMP}, we can always construct an SS-MAMP using optimal damping \cite{Lei2020MAMPISIT, Lei2020MAMPTIT}: Starting with $t=1$ and $\bf{X}_{1}=\bf{0}$,
\BS\label{Eqn:SS_MAMP}\begin{alignat}{2}
  \bf{r}_t &= \gamma^{\rm ss}_t \left( \bf{X}_{t}\right) = \tilde{\bf{R}}_{t} \bf{\zeta}^\gamma_t,\label{Eqn:MAMP_g_con}\\
   \bf{x}_{t + 1} &= \phi^{\rm ss}_t \left(\bf{R}_t\right) = \tilde{\bf{X}}_{t}   \bf{\zeta}^\phi_{t+1},\label{Eqn:MAMP_f_con}
\end{alignat} 
where   
 \begin{align} 
  \bf{\zeta}^\gamma_t &= \left\{\!\!\!\begin{array}{ll}
             \frac{  (\tilde{\bf{V}}^{\gamma}_{t})^{-1} \bf{1}}{\bf{1}^{\rm T} (\tilde{\bf{V}}^{\gamma}_{t})^{-1}\bf{1}}, &\;\; {\rm if}\;   \tilde{\bf{V}}^{\gamma}_{t} {\rm \;is\; invertible} \vspace{1mm}\\
              {\rm [} (\bf{\zeta}^\gamma_{t-1})^{\rm T} \; 0 {\rm ]}^{\rm T}, & \;\; {\rm otherwise}
        \end{array}\right.,  \label{Eqn:dam_g}  \end{align}\begin{align} 
  \bf{\zeta}^\phi_{t+1} &= \left\{\!\!\!\begin{array}{ll}
             \frac{  (\tilde{\bf{V}}^{\phi}_{t+1})^{-1} \bf{1}}{\bf{1}^{\rm T} (\tilde{\bf{V}}^{\phi}_{t+1})^{-1}\bf{1}}, &\;\; {\rm if}\;  \tilde{\bf{V}}^{\phi}_{t+1} {\rm \;is\; invertible} \vspace{1mm}\\
              {\rm [} (\bf{\zeta}^\phi_t)^{\rm T}\; 0 {\rm ]}^{\rm T}, & \;\; {\rm otherwise}
        \end{array}\right., \label{Eqn:dam_f}  
\end{align}\ES   
$ \tilde{\bf{V}}^{\gamma}_{t} =  \tfrac{1}{N}\langle \,  \tilde{\bf{R}}_{t} -\bf{X},\;  \tilde{\bf{R}}_{t} -\bf{X} \,  \rangle$, $\tilde{\bf{V}}^{\phi}_{t+1} = \tfrac{1}{N}\langle \,  \tilde{\bf{X}}_{t+1} -\bf{X},\;   \tilde{\bf{X}}_{t+1} -\bf{X}   \rangle$,  and $(\tilde{\bf{V}}^{\gamma}_{t})^{-1}$ and $(\tilde{\bf{V}}^{\phi}_{t+1})^{-1}$ are the inverses of $\tilde{\bf{V}}^{\gamma}_{t}$ and $\tilde{\bf{V}}^{\phi}_{t+1}$, respectively.   In addition, the following properties hold for the SS-MAMP in \eqref{Eqn:SS_MAMP}.
\begin{enumerate}[(a)] 
    \item Damping preserves the orthogonality, i.e., the  orthogonality in \eqref{Eqn:Orth_MAMP}  holds for the SS-MAMP  in \eqref{Eqn:SS_MAMP}.
    That is  the SS-MAMP in \eqref{Eqn:SS_MAMP} is an MAMP. Following Lemma \ref{Lem:IIDG_MIP}, for the SS-MAMP in \eqref{Eqn:SS_MAMP}, we have: $\forall 1\!\le \!t'\!\leq\! t$, \vspace{-1mm}
    \BS\label{Eqn:IIDG_SS_MAMP}  \begin{align}
    \!\!\!\!\!\!\!\!v_{t,t'}^{\gamma} &\!\!\overset{\rm a.s.}{=}\! \!\tfrac{1}{N}\langle  \gamma_t^{\rm ss}\big(\bf{X}\!\!+\!\!\bf{N}_{t}^{\phi}\big)\!-\!\bf{x},\;  \gamma^{\rm ss}_{t'}\big(\bf{X}\!\!+\!\!\bf{N}_{t'}^{\phi}\big)\!-\!\bf{x} \rangle,\\ 
   \!\!\!\!\!\! v_{t+1,{t'}\!+1}^{\phi} &\!\!\overset{\rm a.s.}{=} \!\!  \tfrac{1}{N}\langle \phi^{\rm ss}_t\big(\bf{X}\!\!+\!\!\bf{N}_{t}^{\gamma} \big)\!-\!\bf{x},\; \phi^{\rm ss}_{t'}\big(\bf{X}\!\!+\!\!\bf{N}_{t'}^{\gamma} \big)\!-\!\bf{x}\rangle,  
    \end{align}\ES 
    where $\bf{N}_{t}^{\gamma}=[\bf{n}_{1}^{\gamma}\dots\, \bf{n}_{t}^{\gamma}]$ and  $\bf{N}_{t}^{\phi}=[\bf{n}_{1}^{\phi} \dots\, \bf{n}_{t}^{\phi}]$ are  column-wise IID Gaussian, row-wise joint-Gaussian and independent of $\bf{x}$. In addition, $ \tfrac{1}{N} \langle \bf{N}_{t}^{\gamma}, \bf{N}_{t}^{\gamma} \rangle =\bf{V}_{t}^{\gamma}$ and $ \tfrac{1}{N}\langle \bf{N}_{t}^{\phi}, \bf{N}_{t}^{\phi} \rangle =\bf{V}_{t}^{\phi}$. In detail, $\bf{n}_t^{\phi}\!\!\sim\!\mathcal{CN}(\bf{0},v_{t,t}^{\phi}\bf{I})$ with ${\rm E}\{\bf{n}_t^{\phi}(\bf{n}_{t'}^{\phi})^{\rm H}\}\!=\!v_{t,t'}^{\phi}\bf{I}$ and $\bf{n}_t^{\gamma}\!\!\sim\!\mathcal{CN}(\bf{0},v_{t,t}^{\gamma}\bf{I})$ with ${\rm E}\{\bf{n}_t^{\gamma}(\bf{n}_{t'}^{\gamma})^{\rm H}\}\!=\!v_{t,t'}^{\gamma}\bf{I}$.    
    
    \item Following Lemma \ref{Lem:L-banded_seq}, the covariance matrices $\bf{V}^{\gamma}_t$ and $\bf{V}^{\phi}_t$ of the SS-MAMP in \eqref{Eqn:SS_MAMP} are L-banded and are determined by sequences $  \bf{v}_t^{\gamma} =[v^{\gamma}_{1}\dots\,  v^{\gamma}_{t}]^{\rm T}$ and $\bf{v}_t^{\phi} =[v^{\phi}_{1}\dots\,  v^{\phi}_{t}]^{\rm T}$, respectively. In detail, 
        \BS\label{Eqn:damped_v}\begin{align}
          \!\!\!v^{\gamma}_{t} &= \left\{\!\!\!\begin{array}{ll}
                  \frac{1}{\bf{1}^{\rm T} (\tilde{\bf{V}}^{\gamma}_{t})^{-1}\bf{1}}, &\;\; {\rm if}\;  \tilde{\bf{V}}^{\gamma}_{t} {\rm \;is\; invertible} \vspace{1mm}\\
                 v^{{\gamma}}_{t-1}, & \;\; {\rm otherwise}
            \end{array}\right.,\label{Eqn:damped_v_g}\\
         \!\!\!  v^{\phi}_{t+1} &= \left\{\!\!\!\begin{array}{ll}
                  \frac{1}{\bf{1}^{\rm T} (\tilde{\bf{V}}^{\phi}_{t+1})^{-1}\bf{1}}, &\;\; {\rm if}\;  \tilde{\bf{V}}^{\gamma}_{t} {\rm \;is\; invertible} \vspace{1mm}\\
                 v^{{\phi}}_{t}, & \;\; {\rm otherwise}
            \end{array}\right..\label{Eqn:damped_v_f}
        \end{align}\ES    
     Following Lemma \ref{Lem:converge}, the state evolution of the SS-MAMP in \eqref{Eqn:SS_MAMP} converges to a fixed point, i.e., 
    \BE\label{Eqn:V_convg} 
        \lim_{t\to\infty} \;v^{\gamma}_{t} \overset{\rm a.s.}{=} v^{\gamma}_*,\qquad
        \lim_{t\to\infty} \; v^{\phi}_{t} \overset{\rm a.s.}{=} v^{\gamma}_*.
    \EE
    
    \item Following Lemma \ref{Lem:damp_useless}, damping is useless (i.e., has no MSE improvement) for the SS-MAMP in \eqref{Eqn:SS_MAMP}. In other words,  the SS-MAMP in \eqref{Eqn:SS_MAMP} obtains the lowest MSE among all the damped MAMP given $\{{\gamma}_t\}$ and $\{{\phi}_t\}$.

    \item The MSE of the SS-MAMP in \eqref{Eqn:SS_MAMP}  is not worse than that of the original MAMP in \eqref{Eqn:org_MAMP}. 
     
\end{enumerate}
\end{theorem}


\begin{remark}
 In Theorem \ref{The:SS-MAMP_con}, if $\tilde{\bf{V}}_t^\gamma$ (or $\tilde{\bf{V}}_{t+1}^\phi$) is singular, $\bf{r}_{t}=\bf{r}_{t-1}$ (or $\bf{x}_{t+1}=\bf{x}_t$).  In this case, we will not allow $\bf{r}_{t}$ (or $\bf{x}_{t+1}$) to join the damping in the next iterations. 
\end{remark}

\section{Conclusion} 
This paper proposes a sufficient statistic memory AMP (SS-MAMP) to solve the convergence problem of the AMP-type algorithms in principle. We show that the covariance matrices of SS-MAMP are L-banded and convergent. Given an arbitrary MAMP, we construct an SS-MAMP by optimal damping. The dynamics of the constructed SS-MAMP can be rigorously described by state evolution. 

In \cite{Lei2022SSMAMP}, using the sufficient statistic technique, we reveal two interesting properties of OAMP/VAMP: 1) the covariance matrices are L-banded and are convergent, and 2) damping and memory are useless. In \cite{Lei2022SSMAMP}, we also prove that SS-MAMP has a faster convergence speed than the MAMP in \cite{Lei2020MAMPISIT, Lei2020MAMPTIT}. 

\appendices 
\section{Proof of Lemma \texorpdfstring{\ref{Lem:SS_Est}}{TEXT}}\label{APP:SS_Est}
\subsection{Proof of \texorpdfstring{\eqref{Eqn:ss_v}}{TEXT}}
     Since $\bf{x}_t = x\bf{1} + \bf{n}_t$ and $\bf{n}_t\sim \mathcal{CN}(\bf{0},\bf{V}_t)$, we have
    \BS\label{Eqn:ps}\begin{align}
     \!\!\! p(x|x_t) & =   \frac{p(x)p(x_t|x)}{p(x_t)} \\ 
                   &\propto   p(x) e^{-v_t^{-1} \|x\|^2 \,+\,  v_t^{-1}x_t^{\rm H}x \,+\,   v_t^{-1}x^{\rm H}x_t}, \\
      \!\!\!  p(x|\bf{x}_t) & = \frac{p(x)p(\bf{x}_t|x)}{p(\bf{x}_t)}\\  
                    &  \propto p(x) e^{-\bf{1}^{\rm T}\bf{V}_t^{-1}\bf{1} \|x\|^2 \,+\,  \bf{x}_t^{\rm H}\bf{V}_t^{-1}\bf{1} x \,+\,   x^{\rm H}\bf{1}^{\rm T}\bf{V}_t^{-1} \bf{x}_t}.
    \end{align}\ES
    From \eqref{Eqn:ps}, for any $\bf{x}_t$, $p(x|\bf{x}_t) =  p(x|x_t)$ if and only if
     \BS\begin{align}
        \bf{1}^{\rm T} \bf{V}_t^{-1} &= [ \bf{V}_t^{-1}\bf{1}]^{\rm T}  = \big[0, \dots, 0, v_{t,t}^{-1}\big],\\ \bf{1}^{\rm T}\bf{V}_t^{-1}\bf{1} &= v_{t,t}^{-1},
    \end{align}\ES
    which is equivalent to  $\bf{V}_t \big[0, \dots, 0, 1\big]^{\rm T} =  v_{t,t}\bf{1}$, i.e., 
    \BE
        v_{t,i}=v_{i,t}=v_{t,t}, \quad \forall 1\le i\le t,
    \EE 
 
 \subsection{Proof  of \texorpdfstring{\eqref{Eqn:useless_dam}}{TEXT}}
  Since $\bf{x}_t = x\bf{1}_{t\times 1}  + \bf{n}_t$ and  $\bf{n}_t$ is zero mean with covariance matrix $\bf{V}_t$, we have
    \BE
       \tfrac{1}{N}\|\bf{x}^{\rm T}_t  \bf{\zeta}_{t} - x \|^2 =  \bf{\zeta}_{t}^{\rm T} \bf{V}_t \bf{\zeta}_{t} 
   \EE
    The damping optimization is equivalent to solve the following quadratic programming problem.
\BE
     \min_{\bf{\zeta}_t} \quad    \tfrac{1}{2} \bf{\zeta}_t^{\rm T} \bf{V}_t \bf{\zeta}_t \qquad
     {\rm s.t.} \quad \bf{1}^{\rm T} \bf{\zeta}_t =1.
\EE 
Since $\bf{V}_t$ is positive semi-definite, it is a convex function with respect to $\bf{\zeta}_t$. We write the Lagrangian function as
    \BE
    {\cal L}(\bf{\zeta}_t, c) =  \tfrac{1}{2} \bf{\zeta}_t^{\rm T} \bf{V}_t \bf{\zeta}_t + c (1-\bf{1}^{\rm T}\bf{\zeta}_t). 
    \EE
    The optimal $\bf{\zeta}_t$ are the solutions of the following equations: 
    \BE
        \nabla_{\!\!\bf{\zeta}_t} {\cal L}(\bf{\zeta}_t, c)  = \bf{0},\qquad
        \partial {\cal L}(\bf{\zeta}_t, c)/\partial c =0.
    \EE 
     That is,  
    \BE\label{Eqn:zeta_opt_cond} 
         \bf{V}_t \bf{\zeta}_t    =  c  \bf{1},\qquad
        \bf{1}^{\rm T}\bf{\zeta}_t  =1.
    \EE
    Since $ v^{\gamma}_{t,1} = v^{\gamma}_{t,2} \dots =  v^{\gamma}_{t,t}$, it is easy to verify that $\{\bf{\zeta}_t = [0\dots\, 0, 1]^{\rm T}, c=v^{\gamma}_{t,t}\}$ is a solution of \eqref{Eqn:zeta_opt_cond}. Hence, $\bf{\zeta}_{t}=[0 \dots 0, 1]$ minimizes the MSE, i.e.,
    \BE
     [0\dots 0\;1]^{\rm T} = \argmin_{\bf{\zeta}_{t}:\bf{\zeta}_{t}^{\rm T}\bf{1}=1} \tfrac{1}{N}\|\bf{\zeta}_{t}^{\rm T}\bf{x}_t -{x} \|^2.
    \EE

\end{document}